\documentstyle[11pt]{nature} \title{
Telling the tale of the first stars} 
\mainauthor{} \author{Timothy C. Beers \affiliation {Department of Physics \&
Astronomy, Michigan State University, E. Lansing, MI 48824, USA}}

\begin{document} \summary{~} \maketitle

\raisebox{13cm}[-13cm]{\emph{Nature} {\bf 422} (2003),	(issue 24 April)}

{\it HE~0107--5240 is a star in more than one sense of the word. Chemically, it is the
most primitive object yet discovered, and it is at the centre of debate about
the origins of the first elements in the Universe.}  

Late last year, discovery of the most iron-deficient star yet identified,
HE~0107--5240, was announced. This star has a measured abundance of iron less
than 1/200,000 that of the Sun\cite{paper1}. Its significance is that it seems to be a relic
from the early Universe, and astronomers are now busy considering how to
interpret it. 

In this issue, three groups -- Bonifacio {\it et al.}\cite{paper2}, Umeda and
Nomoto\cite{paper3}, and Schneider {\it et al.}\cite{paper4} -- present various
interpretations of HE~0107--5240. Each of these contributions centres on whether
this star exhibits properties that might reveal the likely range of mass that
should be associated with the so-called population III stars -- objects that are
presumed to have formed shortly after the Big Bang, and which are thought to
have produced the first elements heavier than H, He, and Li, as well as the
`first light' in the Universe. Population II stars are objects that formed after
population III, and which incorporated the metals created by this previous
generation.  Our Sun, and other (younger) metal-rich stars in the Galactic disk,
are referred to as population I objects. 

To the astronomer, metals include all elements heavier than He, and they are
thought to be produced only by nuclear reactions that take place during the
lifetimes, or at the deaths, of stars. Stars such as our Sun have inherited the
net production of metals by all of the previous generations that lived (and
died) before it. Stars of the lowest observed abundance of heavy elements, such
as HE~0107--5240, must therefore have been born before other stars, because the
gas clouds from which they formed had only the slightest traces of these heavy
elements. So, regardless of the outcome of debate about the nature of the very
first stars, HE~0107--5240 remains the chemically most primitive object yet
discovered, and is a crucial `laboratory' for tests of the origins of the first
elements in the Universe.

Despite the apparent simplicity of the composition of HE~0107--5240, which to
date has yielded detections of nine elements (H, C, N, Na, Mg, Ca, Ti,
Fe, and Ni), the new papers$^{\rm 2-4}$ present a dizzying array of possible
explanations for their origin. By comparison, other extremely low-metallicity
stars, with Fe abundances near 1/1,000 of the solar level, such as
CS~22892--052 (ref. 5) and CS~31082--001 (refs 6, 7), show evidence of roughly
40--60 individual elements, most arising from the so-called rapid
neutron-capture process, which accounts for the production of roughly half of
all elements heavier than Fe. Beyond its Fe deficiency, the singular feature of
HE~0107--5240 is that its measured abundance of C, relative to Fe, is about
10,000 times the observed ratio of these elements in the Sun, the largest such
`over-abundance' ratio ever seen. The N abundance ratio is also highly enhanced,
though only by a factor of 200. The other detected elements exhibit ratios
similar to those in previously identified metal-deficient stars.

Explanations put forward for the composition of HE~0107--5240 fall into three
main categories. First, that it is indeed a low-mass, population III star that
formed out of gas of zero-metallicity, and has had its present surface
abundances altered. Possible mechanisms for changing the observed atmospheric
abundances of HE~0107--5240 include internal mixing of elements produced by
nuclear burning in the star itself, and the acquisition of metals produced by
later generations of stars during its passage through an already enriched
interstellar medium\cite{paper8} or -- as now seems more likely -- even in the
cloud of its birth, from the contribution of material from later
supernovae\cite{paper9}.

The second possible explanation is that it is a low-mass population II star, of
an extreme form, and that the observed elemental abundances directly reflect the
yields of species from supernova explosions of massive (more than 200 times the
solar mass) population III stars that were incorporated into the gas from which
HE~0107--5240 formed.

Third, as above, except that the observed abundances reflect the yields of one
or more supernova explosions of stars of population III stars of `normal' mass
(20--25 solar masses) that were present shortly before its birth. 

An important point here is that each of these options is consistent with the
possibility that HE~0107--5240 has `locked up' the elemental clues required for
understanding the nature of population III. Astronomers have long debated the
form of the so-called `initial mass function' of these objects, although recent
theory has favoured the notion that it was dominated by stars of between several hundred
and one thousand times the mass of the Sun. Such massive stars have extremely
short lifetimes, hence the elements they created provide one of the few
lingering pieces of evidence that can be used to infer their properties.

Detailed observations of HE~0107--5240, now underway, should help to
discriminate between the three options. On page 834, for instance, Bonifacio
{\it et al}\cite{paper2} propose that measurement of the element O may hold one
key. They argue that this element should be detectable in the near-ultraviolet
region of the spectrum of HE~0107--5240. If it turns out to show a ratio, with
respect to Fe, of more than about 1,000, this may be associated with production
from high-mass, zero-metallicity stars. A value of less than 1,000 would suggest
an origin from zero- (or low-) metallicity, lower-mass stars. 

Bonifacio {\it et al.} also suggest that the presently observed metals in
HE~0107--5240 might have arisen from the yields of two supernovae explosions of
zero-metallicity progenitors with quite different masses-- the high-mass
progenitor producing the light elements now observed in the star (but little or
none of the Fe), and the low-mass progenitor contributing the elements Mg and
heavier (including Fe). Similar ideas, calling for a combination of supernovae
to explain the observed elemental abundances of metal-poor stars, have already
been proposed\cite{paper10}. 

Umeda and Nomoto (page 871)\cite{paper3} argue that the observed abundance
pattern in HE~0107--5240, including the large abundances of C and N, can be
explained by the explosions of 20-130-solar-mass progenitors that underwent
substantial `mixing and fallback.' According to this view, most of the heavy
elements (including Fe) that were created in the central regions of explosions
were never ejected, but fell into the maw of a black hole that formed at the
death of the progenitor. The lighter elements, such as C and N, which formed by
pre-explosion nucleosynthesis in the progenitor, were ejected into the
interstellar medium. Umeda and Nomoto also note that similar conditions might
arise from aspherical supernovae explosions, where jets of explosive
nucleosynthesis might have been produced by a rapidly rotating progenitor star.
These ideas are supported by the existence of other C-, N- (and O-) rich,
extremely metal-poor stars, which have abundance patterns that are in some ways
like that of HE~0107--5240 and could be explained by a similar mechanism.
Furthermore, the sort of low-luminosity supernova explosions envisaged by Umeda
and Nomoto have already been observed -- examples include SN1997D and
SN1999br.

Both Umeda and Nomoto\cite{paper3} and Schneider {\it et al.}\cite{paper4} (page
869) point out that the great abundance of C, N (and presumably O) in
HE~0107--5240 may indicate that the gas cloud from which it formed was already
quite metal-rich (even though the Fe abundance is low, the total abundance of
metals is dominated by the lighter elements). That would allow efficient
formation of low-mass stars from the `cooling channels' supplied by the lighter
elements. But these authors also note that, even if the abundances of C, N and O
in the cloud from which HE~0107--5240 formed were quite low, and were boosted
only later from internal mixing processes, gas with metal abundance as low as
that inferred from the Fe alone could still fractionate to form low-mass stars
due to cooling from dust grains -- if indeed such grains could have formed at
sufficiently early times.

Schneider {\it et al.}\cite{paper4} also argue that if the first population III
stars had masses of 200--220 times the solar mass, their explosions might
account for the abundances of the heavier elements in HE~0107--5240, though not
the lighter ones. Schneider {\it et al.} agree with Umeda and Nomoto that
another possibility is the explosion of progenitors with masses 20--25 times
that of the Sun, and point out that improved upper limits on the abundance of
Zn, or the presence (or lack) of elements created in the rapid neutron-capture
process in HE~0107--5240, may be able to discriminate between the appropriate
mass range of the progenitor object. From the data already in hand, however,
Christlieb {\it et al.}\cite{paper1} have argued that if the neutron-capture
elements were indeed greatly enhanced, Ba and Sr might be expected to be
detected in this star. But they are not. 

Although we are left with a frustrating variety of possibilities, HE~0107--5240
should allow us to address many questions about the formation and evolution of
the first generations of stars in the Universe. One line of attack, not
explicitly mentioned in these three papers, is analysis of the metallicity
distribution in stars with the lowest abundances of heavy elements\cite{paper9}.
If they apply in general, several of the explanations offered for the formation
of a star with the observed properties of HE~0107--5240 will produce stars with
characteristic metallicities -- which, given a large enough sample, might be
detected as deviations from a continuous distribution of stellar metallicities.
Hence, numerous additional stars with extremely low Fe abundances will need to
be discovered to fully `tell the tale' of early star formation and the creation
of the first metals in the Universe.

\end{document}